\documentclass[12pt]{article}
\usepackage{amsmath,bm}

\renewcommand{\baselinestretch}{1.4}
\def\be{\begin{equation}}
\def\ee{\end{equation}}
\def\bear{\begin{eqnarray}}
\def\eear{\end{eqnarray}}

\def\bS{\mathbf{S}}

\def\Vol{\mathrm{Vol}}

\def\bi{\bibitem}

\begin{document}

\begin{titlepage}

\begin{flushright}
hep-th/0302086\\
NSF-ITP-03-13
\end{flushright}
\vfil

\begin{center}
{\huge The Sound of M-Theory}\\
%\vspace{3mm}
%{\huge Notes on Sound Waves} \\
\end{center}

\vfil
\begin{center}
{\large Christopher P. Herzog}\\
\vspace{1mm}
Kavli Institute for Theoretical Physics,\\
University of California, Santa Barbara, CA  93106, USA\\
{\tt herzog@kitp.ucsb.edu}\\
\vspace{3mm}
\end{center}

\vfil

\begin{center}

{\large Abstract}
\end{center}

\noindent
We consider sound
propagation on M5- and M2-branes in the hydrodynamic limit.
In particular, we look at the low energy description
of a stack
of $N$ M-branes
at finite temperature.
At low energy, the M-branes are well described, via
the AdS/CFT correspondence, in terms of classical
solutions to the eleven dimensional supergravity equations of motion.
From this gravitational description, 
we calculate Lorentzian signature two-point functions
of the stress-energy tensor  on these M-branes in the
long-distance, low-frequency limit, i.e. the
hydrodynamic limit.
The poles in these Green's functions
show evidence for sound propagation in the field theory
living on the M-branes. 
\vfil
\begin{flushleft}
February 2003
\end{flushleft}
\vfil
\end{titlepage}
\newpage
\renewcommand{\baselinestretch}{1.1}  %looks better

%%%%%%%%%%%%%%%%%%%%%%%%%%%%%%%%%%%%%%%%%%%%%
%% include the next line for double spacing %%
%%%%%%%%%%%%%%%%%%%%%%%%%%%%%%%%%%%%%%%%%%%%%%
%\renewcommand{\baselinestretch}{2}
\renewcommand{\arraystretch}{1.5}

\section{Introduction}

The interacting, superconformal field theories living
on a stack of $N$ M2- or M5-branes are not well understood.
An improved understanding of these M-branes should lead
eventually to a better understanding of M-theory itself, 
a theory that encompasses all the different superstring 
theories and is one of the best hopes for a quantum theory 
of gravity.  
While the full M-brane theories remain mysterious, the
low energy, large $N$ behavior is conjectured to be
described well, via the
AdS/CFT correspondence \cite{jthroat, EW, GKP}, 
by certain classical solutions to eleven dimensional
supergravity equations of motion.  
Recent
work on AdS/CFT correspondence by 
Son, Starinets, Policastro, and the author
\cite{SS, PSS, PSSsound, HS} provides a 
prescription for writing Lorentzian signature
correlators for these types of theories.
In \cite{mthydro}, this prescription was
used to calculate viscosities and diffusion
constants for M-brane theories in this low
energy limit, thus generalizing the work of
\cite{PSS} for D3-branes.  This paper
completes the program started in \cite{mthydro}
by considering sound waves on M2- and M5-branes,
 and also generalizes similar work
by Son, Starinets, and Policastro \cite{PSSsound}
for D3-branes.

The behavior of any field theory perturbed a small
amount away from thermal equilibrium is expected
to be described well by fluid mechanics
\cite{Landau}.
In particular, in this limit the size
of the fluctuations are small compared to the 
temperature.  Moreover, fluid mechanics 
supports sound waves.  Thus, we expect
the finite temperature field theory living
on a stack of M-branes to conduct sound.

The existence of sound waves has strong 
implications for the two-point function of
the stress-energy tensor.  In particular,
we expect to see poles in some
of the components of 
this two-point function
from which we can extract the speed of sound
and a damping term.  

Using the techniques described in \cite{SS, HS}, we
calculate the the relevant components of the 
two-point function of the stress-energy tensor
in these M-brane theories using
their gravity duals.\footnote{
For a different perspective on the 
prescription for writing 
field theory, Lorentzian
signature Green's functions
from gravity using AdS/CFT, 
see \cite{ST}.
}
 We find 
a sound wave pole exactly of the form predicted by 
fluid mechanics:
\be
\omega  = u_s q - i \Gamma q^2 + 
{\mathcal O}(q^3) \ ,
\ee
where $\omega$ is the frequency and $q$
is the wave vector.  
Moreover, we will see that
the speed of sound
$u_s$ and the damping $\Gamma$ take exactly
the values they should.  In particular,
we find that 
$u_s = 1/\sqrt{d}$ where $d$ is the number
of spatial dimensions of the theory.   Also,
$\Gamma$ is related in a precise way to the 
viscosity calculated in \cite{mthydro} from
the non-propagating shear modes. 

There were two main results of \cite{mthydro}.  The first
was that finite temperature AdS/CFT correspondence can
be used to describe M-brane theories in the hydrodynamic
limit.  Indeed, the AdS/CFT description passed some
nontrivial consistency checks, among them two 
independent derivations of the viscosity.
The second result was more evidence in support of the
Lorentzian signature prescription 
proposed in \cite{SS, HS}
for calculating
field theory correlators from gravity. 

The two main results of this paper are largely the 
same.  
With sound waves and more internal consistency checks, among
them a third independent calculation of the viscosity, 
we find additional reason to believe that
AdS/CFT describes well the hydrodynamic limit of 
M-brane theories.  This paper also provides further
support for the Lorentzian signature prescription of
\cite{SS, HS}.

We begin in section 2 by reviewing shear modes and sound
waves in fluid mechanics and the constraints they
place on the two-point function of the stress-energy
tensor.  In section 3, we discuss the field theory
Ward identities.
We show how these identities constrain the components
of the stress-energy two-point function.  In particular,
the identities essentially 
constrain the infrared behavior of the Green's
functions up to the location of the sound wave pole and
a thermal factor.

Section 4 contains the body of the paper, the calculations
of the stress-energy two-point functions from gravity.
These calculations produce the correct sound wave pole
and match precisely the Green's functions obtained
from the Ward identities.

\section{Sound Waves and the Hydrodynamic Limit}

In the hydrodynamic limit, small perturbations of the
stress-energy tensor from thermal equilibrium correspond
to sound and shear modes.  We review  
the differential equations governing these modes.
We are interested in the thermal field theories living
on M2- and M5-branes.  Thus the stress tensor will be
respectively $2+1$ or $5+1$ dimensional.

One of the differential equations is the conservation condition 
$\partial_\mu T^{\mu\nu}=0$.  The second equation
is the linearized form of the stress-energy tensor, which
describes how small, slowly varying 
velocity gradients produce extra
stresses in the fluid because 
of viscosity \cite{Landau}:
%\begin{subequations}
\begin{eqnarray}
T^{ij} = P \delta^{ij} - \frac{\eta}{\epsilon+P}
\left( \partial_i T^{0j} + \partial_j T^{0i} 
- \frac{2}{d} \delta^{ij} \partial_k T^{0k} \right) \ 
\end{eqnarray}
%\end{subequations}
where $P = \langle T^{ii} \rangle$ 
is the pressure, $\epsilon = \langle T^{00} \rangle$ is
the energy density, and $\eta$ is the viscosity.
The dimension $d$ is 2 for the M2-branes and 5 for the M5-branes, and
$i$, $j$, and $k$ run only over the spatial indices.  To analyze
these equations, it is helpful to split $T^{0i}$ into 
longitudinal and transverse parts
(see for example \cite{KadMart}):
\be
T^{0i} = g_i^{(t)} + g_i^{(\ell)} \ .
\ee  
In more abstract language,
we treat $T^{0i}$ as a one-form that lives in Euclidean
$d$-dimensional space, and we express the form as a sum of a closed
(or longitudinal) one-form $g^{(\ell)}$ 
and a co-closed (or transverse) one-form $g^{(t)}$.  In index notation,
$\partial_i g_i^{(t)} = 0$ and $\partial_i g_j^{(\ell)}
= \partial_j g_i^{(\ell)}$.

For the transverse or shear modes, the conservation condition and
the linearized stress-energy tensor combine to give 
\be
\partial_0 g_i^{(t)} =  \frac{\eta}{\epsilon+P} (\partial_j)^2 
g_i^{(t)} \ .
\ee
%The conservation condition further implies 
%that the shear modes do not cause
%any variation in $T^{00}$.  
In short, the shear modes obey a diffusion law,
and one expects a corresponding diffusion pole
$i\omega = D q^2$ in the Green's functions
where $D = \eta / (\epsilon + P)$. 
These modes were 
analyzed in \cite{mthydro}, where it was discovered
from a gravity calculation that 
$D= 1/4\pi T$ 
for both the M2- and M5-branes to leading
order in $1/N$.

For the longitudinal or sound modes, one finds that
\be
\left(-(\partial_0)^2 + u_s^2 (\partial_i)^2
+\frac{2 \eta (d-1)}{(\epsilon+P)d} (\partial_i)^2 \partial_0 
\right) T^{00} = 0
\ee
along with the subsidiary relation $\partial_0 T^{00} = \partial_i g_i^{(\ell)}$.  We have defined the speed of sound
$u_s^2 \equiv \partial P / \partial \epsilon$.  
For a perturbation of the form 
$T^{00} = \epsilon + a e^{-i\omega t + i q x}$ where 
$a \ll \epsilon$, there is a dispersion relation of the form
\be
\omega = \pm u_s q - \frac{i (d-1)}{d} \frac{\eta}{\epsilon+P} q^2 
+ {\mathcal O}(q^3) \ .
\label{disp}
\ee
The theories we are considering are conformal and so the stress-energy tensor is traceless.  In other words $\epsilon = Pd$ and 
$u_s = 1/\sqrt{d}$.

\section{Ward Identities}

The one- and two-point functions of the stress-energy tensor are 
highly constrained by Ward identities.  
In particular, we will find that the components of the two-point
function relevant to sound propagation are constrained up
to one free parameter $\alpha$.  
In the following, we will 
review the relevant Ward identities and try to motivate
a particular choice of $\alpha$.

In a flat metric,
the Ward identity for $\langle T^{\mu\nu} \rangle$ is 
conservation of energy, $\partial_\mu \langle T^{\mu\nu} \rangle = 0$.
For our thermal field theories, $\langle T^{\mu\nu} \rangle$ is a 
constant with $\langle T^{00}\rangle =\epsilon$ and 
$\langle T^{ij} \rangle = P \delta^{ij}$.

For the two-point function,
\be
iG_F^{\mu\nu\lambda\rho}(q) = 
\int d^{d+1}y \, e^{-iq\cdot y} 
\langle T \left( T^{\mu\nu}(y) T^{\lambda \rho}(0)\right) \rangle \ ,
\ee
the Ward identity takes a more complicated form:
\be
q_\mu \left(G^{\mu\nu\lambda\rho}(q) - \eta^{\nu\lambda}
\langle T^{\mu\rho} \rangle 
-\eta^{\nu \rho} \langle T^{\mu\lambda} \rangle
+\eta^{\mu\nu} \langle T^{\lambda\rho} \rangle \right) = 0 \ .
\label{Wardo}
\ee
The conformal invariance of our field theories leads to
an additional conformal Ward identity:
\be
\eta_{\mu\nu} G^{\mu\nu \lambda \rho}(q) = 2 
\langle T^{\lambda\rho} \rangle \ .
\label{Wardt}
\ee
As mentioned before, the Ward identities do not completely
constrain $G$, 
and we have removed the subscript from the Green's function
so that we may think of $G$ as  
any function that
satisfies these identities (\ref{Wardo}) and (\ref{Wardt}).

%The identities can be used to constrain highly 
%the low-energy behavior of the $G^{\mu\nu\lambda\rho}$ relevant
%to sound propagation.  
We now identify the particular components of 
$G^{\mu\nu\lambda\rho}$ relevant to sound
propagation.
We choose our sound wave to have
wave vector $(\omega, q, 0, 0, \ldots)$.  In this frame,
we are only interested in components of $T^{\mu\nu}$ which
are invariant under rotations in the remaining $d-1$ 
spatial directions perpendicular to the $x_1=x$ axis, i.e.
$T^{tt}$, $T^{tx}$, $T^{xx}$, and $T^{aa}=T^{x^2x^2} + 
\ldots + T^{x^dx^d}$.
Therefore, there are ten independent correlators
$G^{AB}$ where $A$ and $B$ are $tt$, $xx$, $tx$, or $aa$.

There are twelve Ward identities for these correlators, nine of which are linearly independent.  
Thus, there is a one parameter family of Green's functions
$G$ that satisfy these Ward identities.
Let $G_1$ be a particular solution to the identities
such that the complex conjugate $G_1^* \neq G_1$.
As the identities have real
coefficients, $G_1^*$ will
also be a solution, and we can write the most
general solution
as
\be
G = (1+\alpha) G_1 - \alpha  G_1^* \ ,
\label{fformo}
\ee
where $\alpha$ is the free parameter.

We now find a particular solution $G_1$ 
such that $G_1 \neq G_1^*$ with 
a desirable quality: $G_1$ will be
closely related to the retarded Green's
function $G_R$.
We expect the only singularities of the {\it retarded}
Green's functions, $G_R$, to be simple poles at 
$\omega= \pm q/\sqrt{d} - i\Gamma q^2 + {\mathcal O}(q^3)$, 
corresponding to sound propagation; $\Gamma>0$ is the
damping term.  
Let us assume that $G_1$ has the same pole structure
as $G_R$.
With this additional constraint, the correlators become
\be
G_1^{\mu\nu\lambda\rho}(\omega, q) = \frac{P}
{(d\omega^2 - q^2) + i\beta q^2 \omega}
P^{\mu\nu\lambda\rho}(\omega, q)
\ee
where the $P^{\mu\nu\lambda\rho}$ are polynomials in $\omega$ and $q$.
The constant $\beta=2\Gamma d$ is proportional to the damping.
That the $P^{\mu\nu\lambda\rho}$ are polynomials uniquely
constrains them to be
\begin{subequations}
\begin{eqnarray}
%\begin{array}{ll}
P^{tttt} &=& d \left[ \left((d+2)q^2-d\omega^2 \right)
%(1+\alpha \omega)
-i\beta q^2\omega \right] \ , 
\label{Ward1} \\
P^{tttx} &=& \omega q d (d+1) 
%(1+\alpha \omega) \ ,
\\
P^{ttxx} &=& d \left[ (\omega^2 + q^2) 
%(1 + \alpha \omega) 
- i \beta q^2 \omega \right] \ , \\
P^{txtx} &=& (d^2 \omega^2 + q^2) 
%(1 + \alpha \omega)
 - i \beta q^2 \omega \ , \\
P^{txxx} &=& \omega q (1+d) (1
%+ \alpha \omega 
- i \beta \omega) \ , \\
P^{xxxx} &=& \left( (1+2d)\omega^2 -q^2 \right)
% (1+\alpha \omega)
- i\beta (1+d) \omega^3 + i \beta q^2 \omega 
\ , \\
P^{ttaa} &=& d \left[ (d-1)(\omega^2+q^2)
%(1+\alpha \omega)
+2i\beta q^2 \omega \right]
\ , \\
P^{txaa} &=& \omega q (d^2-1)
%(1+\alpha \omega)
+ i \beta (1+d) \omega^2 q 
\ , \\
P^{xxaa} &=& (d-1) (\omega^2 + q^2)
%(1+\alpha \omega)
+ i \beta(1+d) \omega^3 + i \beta (1-d) q^2 \omega
\ , \\
P^{aaaa} &=&(d-1) \left((3d-1)\omega^2 +(d-3)q^2 \right)
%(1+\alpha \omega) 
\nonumber \\
&& - i \beta(1+d) \omega^3 + i \beta(5d-3)q^2 \omega \ .
%end{array}{ll}
\label{Wardtwopt}
\end{eqnarray}
\end{subequations}
Because they have the same pole structure $G_1$
and $G_R$ are closely related.  However, it was
shown in \cite{PSSsound} that the two differ by
real contact terms. 
For example $G_R^{tttt} = G_1^{tttt} + \epsilon$.

%Let us return to the difference between the $G$ and the $G_R$.
%The retarded Green's functions $G_R$ are defined as
%\be
%G_R^{\mu\nu\lambda\rho}(q) = 
%-i \int d^{d+1}x \, e^{-iq\cdot x} \theta(x^0)
%[T^{\mu\nu}(x), T^{\lambda\rho(0)] \ .
%\ee
%As was noted in \cite{PSSsound}, for ${\bf q}=0$,
%$G_R^{t\nu\lambda\rho}(\omega, 0) = 0$ because
%of translational invariance.  However,
%putting $q=0$ in the expressions for $G$ above
%(\ref{Wardtwopt}), one sees immediately that
%$G$ does not always vanish.  In short, 
%$G$ differs from $G_R$ by contact terms.
%

To compare with the gravity calculations
to come, we need an expression for
the Feynman Green's function $G_F$. 
As the $G$ in the Ward identities are 
defined by
taking functional derivatives of an
appropriately defined path integral,
we expect that one of these solutions
to the Ward identities
should be identified
with the Feynman Green's functions, $G_F$.
There
must be some $\alpha$ 
in (\ref{fformo})
for which $G=G_F$.
%Because they have the same pole structure,
%$G_1$ and the retarded Green's function $G_R$
%are closely related.
%Indeed, it was shown in \cite{PSSsound} that
%the two differ only by real contact terms.
From the definitions of the various
Green's functions and the 
Kubo-Martin-Schwinger (KMS) 
condition, we know that
$G_F = (1+n) G_R - n G_R^*$, 
up to real contact terms, where
$n = (\exp(\omega/T)-1)^{-1}$.
Thus it seems likely that
$G=G_F$ in (\ref{fformo}) when $\alpha = n$, i.e.
\be
G_F = (1+n) G_1 - n G_1^* \ .
\label{fform}
\ee
There has been some guess work involved in arriving
at this expression for $G_F$, namely in the particular
choice $\alpha=n$.  However, we will see that this
form for $G_F$ matches precisely the gravity
calculations in the following section.

%\begin{equation}
%\begin{array}{lll}
%P^{tttt} = d(d+2)q^2-d^2\omega^2 &
%P^{tttx} = d(d+1) \omega q &
%P^{ttxx} = d(q^2 + \omega^2) \\
%%
%P^{txtx} = q^2 + d^2 \omega^2 &
%P^{txxx} = (d+1) \omega q &
%P^{xxxx} = -q^2 + (2d+1)\omega^2 \\
%%
%P^{ttaa} = d(d-1)(\omega^2 + q^2) &
%P^{txaa} = (d^2-1) \omega q &
%P^{xxaa} = (d-1) (\omega^2+q^2) \\
%%
%P^{aaaa} = (d-1) ((3d-1) \omega^2 + (d-3) q^2 ) 
%%
%\end{array}
%\label{Wardtwopt}
%\end{equation}

\section{Two-Point Functions from Gravity}

In the following, we check that a gravitational
calculation produces stress-energy correlators
that obey the Ward identities.  Indeed,
we will find that the M2- and M5-brane 
correlators calculated from gravity have
precisely the form (\ref{fform}).

\subsection{M2-branes}

The background metric is 
\be
ds^2 = \left( \frac{2\pi T R}{3u}\right)^2 \left[
-f(u) dt^2 + dx^2 + dy^2 \right] +
\frac{R^2}{4 u^2 f(u)} du^2
\label{bmet}
\ee
where $f(u) = 1-u^3$,
and $R$ is the radius of curvature
of AdS.  
We expect the gravitational description of the
M-branes to
be valid in the limit where $R$ is large
compared to the eleven dimensional Planck
length $\ell_P$.
The number $N$ of M2-branes is related
to $\ell_P$ and $R$ 
through Dirac quantization, 
$R/\ell_P \sim N^{1/6}$ \cite{Kleb}.
Thus, we are working in
the large $N$ limit, and there could 
in principle be ${\mathcal O}(1/N)$ 
corrections to our results.

To study the two-point functions
of the stress-energy tensor,
we introduce perturbations to
this background metric of the form
$g_{\mu\nu} = g^{(0)}_{\mu\nu}+ h_{\mu\nu}$.
We choose a gauge where $h_{u\mu}=0$.  
We consider waves that move in the $x$ direction.
Such waves break into two categories, depending on
whether the metric perturbations are odd or even
under $y \to -y$.  The odd perturbations correspond
to diffusive shear waves and were treated in 
\cite{mthydro}.  The even perturbations correspond
to sound waves and will be treated here.
For sound waves, the nonzero elements of $h_{\mu\nu}$
are $h_{xx}$, $h_{yy}$, $h_{tt}$, and $h_{xt}$.

We decompose the metric perturbations into their
Fourier components
\be
h_{\mu\nu}(t,x, u) = \int \frac{d\omega dq}{(2\pi)^2} 
e^{-i\omega t + iqx} h_{\mu\nu} (\omega, q, u) \ .
\label{Fcomp}
\ee
We also introduce the dimensionless energy and momentum
\[
\omega_2 = \frac{3 \omega}{2\pi T} \; \; \; ; \; \; \; 
q_2 = \frac{3 q}{2\pi T} \ .
\]

To first order in the $h_{\mu\nu}$, the Einstein equations
are
\begin{subequations}
\be
H_{tt}'' + \frac{1}{2u} \left( -\frac{9}{f}+7 \right) H_{tt}'
+H_s'' + \frac{1}{2u} \left( -\frac{3}{f} + 1 \right) H_s' = 0 \ ,
\ee
\be
H_{tt}'' + \frac{1}{2u} \left( -\frac{9}{f} + 3 \right) H_{tt}' 
+\frac{1}{2u} \left( -\frac{3}{f} + 1 \right) H_s' - \frac{q_2^2}{4f}H_{tt}
+ \frac{\omega_2^2}{4f^2} H_s + \frac{\omega_2 q_2}{2f^2} H_{xt} = 0 \ ,
\ee
\be
H_s'' - \frac{2}{u} H_{tt}' + \frac{1}{u} \left( - \frac{3}{f} -1 \right) H_s'
- \frac{q_2^2}{4f}H_{tt} + \frac{\omega_2^2}{4f^2} H_s - \frac{q_2^2}{2f}H_{yy}
+ \frac{\omega_2 q_2}{2f^2} H_{xt} = 0 \ ,
\ee
\be
H_{yy}'' - \frac{1}{u} H_{tt}' - \frac{1}{u}H_s' -\frac{1}{u}
\left(\frac{3}{f} - 1\right)H_{yy}' + \frac{1}{4f^2}
\left(\omega_2^2 - fq_2^2 \right) H_{yy} = 0 \ ,
\ee
\be
H_{xt}'' - \frac{2}{u}H_{xt}' + \frac{\omega_2 q_2}{4f}H_{yy}=0\ , 
\ee
\be
q_2(-2 f H_{tt}' +3 u^2 H_{tt})- 2q_2 f H_{yy}' + 2\omega_2 H_{xt}' = 0 \ ,
\ee
\be
\omega_2 (2f H_s' +3 u^2 H_s)+ q_2 (2 f H_{xt}' + 6 u^2 H_{xt}) = 0 \ .
\ee
\end{subequations}
We have defined $H_s = H_{xx} + H_{yy}$ and $H_{tt} = h_t^t$, $H_{xt} = h_t^x$, 
$H_{ii} = h_i^i$.  Here $h_\mu^\nu$ are the Fourier decomposed 
metric perturbations where the indices have been raised with the background
metric $g_{\mu\nu}^{(0)}$ (\ref{bmet}).

The system above is not linearly independent, but can be
reduced to the following four linearly independent
differential equations
\begin{subequations}
\be
H_{tt}'' + \frac{1}{2u} \left( -\frac{9}{f}+7 \right) H_{tt}'
+H_s'' + \frac{1}{2u} \left( -\frac{3}{f} + 1 \right) H_s' = 0 \ ,
\ee
\be
q_2(-2 f H_{tt}' +3 u^2 H_{tt})- 2q_2 f H_{yy}' + 2\omega_2 H_{xt}' = 0 \ ,
\ee
\be
\omega_2 (2f H_s' +3 u^2 H_s)+ q_2 (2 f H_{xt}' + 6 u^2 H_{xt}) = 0 \ ,
\ee
\be
\frac{4}{u} H_{tt}' + \frac{1}{u} \left( \frac{3}{f} + 1 \right) H_s'
+ \frac{q_2^2}{2f}(H_{tt} + H_{yy} )
-\frac{\omega_2^2}{2f^2} H_s - \frac{\omega_2 q_2}{f^2} H_{xt} = 0 \ .
\ee
\end{subequations}
The five integration constants correspond to the choice of 
incoming boundary condition at $u=1$ and the four Dirichlet
boundary conditions at $u=0$.

These systems of differential equations are invariant under
three residual gauge transformations.  These pure
gauge solutions are linear combinations of 
$H^I$, $H^{II}$, and $H^{III}$ whose explicit forms are
\begin{subequations}
\begin{eqnarray}
H^{I}_{xt} &=& -\omega_2 \ ,\\
H^{I}_{xx} &=& 2 q_2 \ ,
\end{eqnarray}
\end{subequations}
\begin{subequations}
\begin{eqnarray}
H^{II}_{tt} &=& 2\omega_2 \ ,\\
H^{II}_{xt} &=& q_2 f \ ,
\end{eqnarray}
\end{subequations}
\begin{subequations}
\begin{eqnarray}
H^{III}_{tt} &=& -2\omega_2^2 \int \frac{u}{f^{3/2}} du - 4\left(\frac{2+u^3}{f^{1/2}}\right) \ , \\
H^{III}_{xt} &=& -\omega_2 q_2 \left( \int \frac{u}{f^{1/2}}du + 
f \int \frac{u}{f^{3/2}} du \right) \ , \\
H^{III}_{xx} &=& 2q_2^2 \int \frac{u}{f^{1/2}} du - 8 f^{1/2} \ , \\
H^{III}_{yy} &=& - 8 f^{1/2} \ .
\end{eqnarray}
\end{subequations}

There is an incoming solution to these differential equations,
where by incoming we mean that the wave fronts at the horizon
are purely incoming.
The incoming solution to linear 
order\footnote{
In \cite{PSSsound}, the authors used the 
quadratic order solution to $H^{inc}$
to compute the correlators and the damping. 
We will see, however, that the 
correct damping appears already at linear order.}
 in $\omega_2$ and $q_2$ is
\begin{subequations}
\begin{eqnarray}
H^{inc}_{tt} &=& {\mathcal O}(\omega_2^2, \omega_2q_2, q_2^2) \ , \\
H^{inc}_s &=& {\mathcal O}(\omega_2^2, \omega_2 q_2, q_2^2) \ , \\
H^{inc}_{xt} &=& (1-u)^{-i\omega_2/6} \left( -\frac{iq_2}{6} f(u) + 
{\mathcal O}(\omega_2^2, \omega_2q_2, q_2^2)  
\right) \ , \\
H^{inc}_{yy} &=& (1-u)^{-i\omega_2/6} 
\left( 1 - \frac{i\omega_2}{6} \ln \frac{1+u+u^2}{3} +
{\mathcal O}(\omega_2^2, \omega_2q_2, q_2^2)  \right) \ .
\end{eqnarray}
\end{subequations}
There is also naturally an outgoing solution at the horizon which
is just the complex conjugate of $H^{inc} = (H^{out})^*$.
The nonperturbative piece $(1-u)^{\pm i\omega_2/6}$ enforces
the incoming or outgoing boundary condition at the 
horizon $u=1$, as discussed in more detail in
\cite{PSS, mthydro}.

Setting the boundary conditions 
is a delicate issue.  In \cite{HS}, it was argued that the
correct boundary conditions for calculating field
theory propagators using AdS/CFT are purely incoming for
positive frequency modes and purely outgoing for
negative frequency modes where positive and negative 
frequency are defined with respect to Kruskal time.
Moreover, the full Penrose diagram 
for this black hole background and a second boundary
were used to reproduce the $2\times 2$ matrix
of Schwinger-Keldysh propagators.  For simplicity,
we will focus on deriving just the Feynman propagator, 
i.e. the first entry in this $2\times 2$
matrix, in what follows. 

Let $H(u)$ be the solution to the system of ODE's which
contains no $H^{out}$ component, i.e.
%The solution to this system of ODE's is thus a linear combination
%of 
\be
H(u) = a H^{inc}(u) + bH^I(u) + cH^{II}(u) + dH^{III}(u) \ ,
\label{bulkgravone}
\ee
where $a$, $b$, $c$, and $d$ depend on the boundary values
of the bulk graviton field $H_{tt}^0$, $H_s^0$, $H_{yy}^0$, and
$H_{xt}^0$ in the limit $u \to 0$.
Ignoring the dependence of the propagators on the 
second boundary, 
the bulk-graviton wave-function $H_K$ with the more complicated
Kruskal time-dependent boundary conditions can be written in
terms of $H(u)$.  In particular, $H_K = (1+n) H(u) - n H(u)^*$
where $n = (\exp(\omega/T)-1)^{-1}$, as discussed in \cite{HS}.
As the $H_K$ can be constructed out of $H(u)$ without
too much difficulty, it is convenient to focus on
$H(u)$.  The fact that $H(u)$ does not rely on the
outgoing piece of the bulk-graviton wave-function means
that it has a causal structure and is thus more closely
connected with the retarded Green's function $G_R$ 
in field theory while $H_K$ is related to the Feynman
propagator.

Solving this system (\ref{bulkgravone})
of equations to linear order in 
$\omega_2$ and $q_2$, one finds that the constants
$a$, $b$, $c$, and $d$ are all proportional to
\be
\left( \left(1 + \frac{i\omega_2}{6} \ln(3) \right) 
(2\omega_2^2 - q_2^2) + \frac{i q_2^2 \omega_2}{3} \right)^{-1} \ .
\ee
 In other words,
the Green's function will have a pole at
\be
\omega = \pm \frac{q}{\sqrt{2}} - \frac{iq^2}{8\pi T} + {\mathcal O}(q^3)
\ee
 corresponding
to sound waves with a speed of $1/\sqrt{2}$, 
as expected.  
Moreover, from \cite{mthydro} we know that
$\eta / (\epsilon+P) = 1/4\pi T$.  Thus
the damping term matches
what is expected from the dispersion relation (\ref{disp}).

\subsection{M5-branes}

The calculations for the M5-brane are very similar.
The background metric is 
\be
ds^2 = \left( \frac{4\pi T R}{3}\right)^2 \frac{1}{u} \left[
-f(u) dt^2 + d\vec x^2 \right] +
\frac{R^2}{u^2 f(u)} du^2
\label{bmet7}
\ee
where $f(u) = 1-u^3$
and $\vec x = (x^1, x^2, x^3, x^4, x^5)$.  
We introduce perturbations to
this background metric of the form
$g_{\mu\nu} = g^{(0)}_{\mu\nu}+ h_{\mu\nu}$.
We choose a gauge where $h_{u\mu}=0$.  
We consider waves that move in the $x^1 \equiv x$ direction.
Such waves break into two categories, depending on
whether the metric perturbations transform as a 
vector or a scalar under the residual rotations
in the directions transverse to $x$ and $t$.
The vector perturbations correspond
to diffusive shear waves and were treated in 
\cite{mthydro}.  The scalar perturbations correspond
to sound waves.
For sound waves, the nonzero elements of $h_{\mu\nu}$
are $h_{xx}$, $h_{tt}$, $h_{xt}$, and 
$h_{yy} \equiv h_{x^2x^2}=h_{x^3x^3}=h_{x^4x^4}=h_{x^5x^5}$.

We again decompose the metric perturbations into their
Fourier components (\ref{Fcomp}).
We also introduce the dimensionless energy and momentum
\[
\omega_5 = \frac{3 \omega}{4\pi T} \; \; \; ; \; \; \; 
q_5 = \frac{3 q}{4\pi T} \ .
\]

To first order in the $h_{\mu\nu}$, the Einstein equations
are
\begin{subequations}
\be
H_{tt}'' - \frac{9u^2}{2f} H_{tt}'
+H_s'' - \frac{3u^2}{2f} H_s' = 0 \ ,
\ee
\be
H_{tt}'' + \frac{1}{u} \left( -\frac{9}{2f} + 2 \right) H_{tt}' 
+\frac{1}{u} \left( -\frac{3}{2f} + 1 \right) H_s' - \frac{q_5^2}{uf}H_{tt}
+ \frac{\omega_5^2}{uf^2} H_s + \frac{2\omega_5 q_5}{uf^2} H_{xt} = 0 \ ,
\ee
\be
H_s'' - \frac{5}{2u} H_{tt}' - \frac{3}{2u} \left(  \frac{2}{f} +1 \right) H_s'
- \frac{q_5^2}{uf}H_{tt} + \frac{\omega_5^2}{uf^2} H_s - \frac{8q_5^2}{uf}H_{yy}
+ \frac{2\omega_5 q_5}{uf^2} H_{xt} = 0 \ ,
\ee
\be
H_{yy}'' - \frac{1}{2u} H_{tt}' - \frac{1}{2u}H_s' +\frac{1}{u}
\left(-\frac{3}{f} + 1\right)H_{yy}' + \frac{1}{uf^2}
\left(\omega_5^2 - fq_5^2 \right) H_{yy} = 0 \ ,
\ee
\be
H_{xt}'' - \frac{2}{u}H_{xt}' + \frac{4\omega_5 q_5}{uf}H_{yy}=0\ , 
\ee
\be
q_5(3u^2 H_{tt} -2 f H_{tt}')- 8q_5 f H_{yy}' + 2\omega_5 H_{xt}' = 0 \ ,
\ee
\be
\omega_5 (2f H_s' +3 u^2 H_s)+ q_5 (2 f H_{xt}' + 6 u^2 H_{xt}) = 0 \ .
\ee
\end{subequations}
We have defined $H_s = H_{xx} + 4H_{yy}$ and $H_{tt} = h_t^t$, $H_{xt} = h_t^x$, 
and so forth as before.  

This system is equivalent to the four linearly
independent equations
\begin{subequations}
\be
H_{tt}'' - \frac{9u^2}{2f} H_{tt}'
+H_s'' - \frac{3u^2}{2f} H_s' = 0 \ ,
\ee 
\be
q_5(3u^2 H_{tt} -2 f H_{tt}')- 8q_5 f H_{yy}' + 2\omega_5 H_{xt}' = 0 \ ,
\ee
\be
\omega_5 (2f H_s' +3 u^2 H_s)+ q_5 (2 f H_{xt}' + 6 u^2 H_{xt}) = 0 \ ,
\ee
\be
\frac{5}{u} H_{tt}' + \frac{1}{u} \left( \frac{3}{f} + 2 \right) H_s'
+ \frac{2q_5^2}{uf}(H_{tt} + 4H_{yy} )
-\frac{2\omega_5^2}{uf^2} H_s - \frac{4\omega_5 q_5}{uf^2} H_{xt} = 0 \ .
\ee
\end{subequations}

These systems of differential equations are invariant under
three residual gauge transformations.  These pure
gauge solutions are linear combinations of 
$H^I$, $H^{II}$, and $H^{III}$ whose explicit forms are
\begin{subequations}
\begin{eqnarray}
H^{I}_{xt} &=& -\omega_5 \ ,\\
H^{I}_{xx} &=& 2 q_5 \ ,
\end{eqnarray}
\end{subequations}
\begin{subequations}
\begin{eqnarray}
H^{II}_{tt} &=& 2\omega_5 \ ,\\
H^{II}_{xt} &=& q_5 f \ ,
\end{eqnarray}
\end{subequations}
\begin{subequations}
\begin{eqnarray}
H^{III}_{tt} &=& -2\omega_5^2 \int \frac{1}{f^{3/2}} du - \left(\frac{1+2u^3}{f^{1/2}}\right) \ , \\
H^{III}_{xt} &=& -\omega_5 q_5 \left( \int \frac{1}{f^{1/2}}du + 
f \int \frac{1}{f^{3/2}} du \right) \ , \\
H^{III}_{xx} &=& 2q_5^2 \int \frac{1}{f^{1/2}} du - f^{1/2} \ , \\
H^{III}_{yy} &=& - f^{1/2} \ .
\end{eqnarray}
\end{subequations}

The incoming solution to linear order in $\omega_2$ and $q_2$ is
\begin{subequations}
\begin{eqnarray}
H^{inc}_{tt} &=& {\mathcal O}(\omega_5^2, \omega_5q_5, q_5^2) \ , \\
H^{inc}_s &=& {\mathcal O}(\omega_5^2, \omega_5 q_5, q_5^2) \ , \\
H^{inc}_{xt} &=& (1-u)^{-i\omega_5/3} \left( -\frac{4iq_5}{3} f(u) + 
{\mathcal O}(\omega_5^2, \omega_5q_5, q_5^2)  
\right) \ , \\
H^{inc}_{yy} &=& (1-u)^{-i\omega_5/3} 
\left( 1 - \frac{i\omega_5}{3} \ln \frac{1+u+u^2}{3} +
{\mathcal O}(\omega_5^2, \omega_5q_5, q_5^2)  \right) \ .
\end{eqnarray}
\end{subequations}

The solution to this system of ODE's 
with pure incoming boundary conditions at the horizon
is thus a linear combination
of 
\be
H(u) = a H^{inc}(u) + bH^I(u) + cH^{II}(u) + dH^{III}(u) \ .
\label{bulkgravtwo}
\ee
Solving this system of equations to linear order in 
$\omega_5$ and $q_5$, one finds that the constants
$a$, $b$, $c$, and $d$ are all proportional to
\be
\left( \left(1 + \frac{i\omega_5}{3} \ln (3) \right) 
(5\omega_5^2 - q_5^2) + \frac{8i \omega_5 q_5^2}{3} \right)^{-1} \ . 
\ee 
In other words,
the Green's function will have a pole at
\be
\omega = \pm \frac{q}{\sqrt{5}} - \frac{i q^2}{5 \pi T} + {\mathcal O}(q^3)
\ee 
corresponding
to sound waves with a speed of $1/\sqrt{5}$, 
as expected.  Moreover, the damping term matches precisely
(\ref{disp}), given that $\eta/ (\epsilon + P) = 1 /4\pi T$
\cite{mthydro}.

\subsection{Two-Point Functions from the Boundary Action}

Although one can see immediately from the constants 
$a$, $b$, $c$, and $d$ in the solution to the bulk graviton
(\ref{bulkgravone}) and (\ref{bulkgravtwo}) that the
two-point functions of the stress-energy tensor will
have poles that are consistent with sound wave propagation,
it is a worthwhile exercise to work out the two-point
functions explicitly using the
AdS/CFT prescription.  One benefit is that we will
be able to check the two-point functions calculated
from the gravity against the formulae we derived from
the Ward identities (\ref{Ward1})--(\ref{Wardtwopt}).

In order to use the AdS/CFT prescription, we need
to find the boundary action for our gravitational
systems.  The full action is
\be
\frac{R^{10-d-1} \Vol(\bS^{10-d-1})}{2 \kappa_{11}^2}
\left( \int d^{d+1}x \, du \,
\sqrt{g} \left( {\mathcal R} - 2\Lambda \right)
+
\int d^{d+1}x \, \sqrt{g^B} (2K + a) \right) \ 
\ee
where $\Lambda$ is a bulk cosmological constant, $a$ is a
boundary cosmological constant that is 
necessary to cancel overall $H$ independent 
divergences in the boundary action, $K$ is the
extrinsic curvature, and $g^B_{ij}$ is the metric
on the boundary $u=0$.
The number $\Vol(\bS^n)$ is the volume of a unit
$n$-dimensional sphere.
For $AdS_4$, $d=2$, $\Lambda=-12/R^2$, and $a=-8$,
while 
for $AdS_7$, $d=5$, $\Lambda=-15/4R^2$, and $a=-5$.  
The gravitational coupling constant $\kappa_{11}$
is related by Dirac quantization to the number of
M2- and M5-branes.  For the M2-branes,
$R^9 \pi^5 =  N^{3/2} \kappa_{11}^2 \sqrt{2}$, while
for the M5-branes, 
$N^3 \kappa_{11}^2 = 2^7 \pi^5 R^9$ \cite{Kleb}.
A more detailed discussion of this bulk
action can be found in 
\cite{LiuTseyt}.

Using the equations of motion, the full
action can be reduced to a boundary term.  
This boundary action for the M2-branes is
\begin{eqnarray}
S_b &=& \frac{P}{8} \int d^3x 
\left[ -8 -8H_{tt} + 4H_{xx} + 4H_{yy} \right. \nonumber \\
&&
-\frac{2}{\epsilon^2} \left( H_{tt} (H_{xx} + H_{yy})
 + H_{xx}H_{yy} + H_{xt}^2 \right)' \nonumber \\
& &
\left. - (H_{xx} + H_{yy})H_{tt} - H_{xx}^2 + 2 H_{xx}H_{yy} - H_{yy}^2
+2H_{tt}^2 - 8 H_{xt}^2 \right] \ ,
\label{bryMt}
\end{eqnarray}
where the constant $P = 8 \sqrt{2} \pi^2 T^3 N^{3/2} / 81$ 
is the pressure.

For the M5-branes, the boundary action is
\begin{eqnarray}
S_b &=& \frac{P}{8} \int d^6x \bigr[
-8 -20H_{tt} +4 H_{xx} + 4H_{aa}  \nonumber \\
&&
-\frac{1}{\epsilon^2} \left(
4H_{tt} (H_{xx} + H_{aa}) +
\frac{3}{2} H_{aa}^2 + 4H_{aa}H_{xx} + 4H_{xt}^2 \right)'
\nonumber \\
&&
\left. -4H_{tt}(H_{xx}+H_{aa}) - H_{xx}^2 + 2H_{aa}H_{xx}+ \frac{1}{2}H_{aa}^2
+ 5H_{tt}^2 - 20 H_{xt}^2 \right]
\label{bryMf}
\end{eqnarray}
where $P = 2^6 \pi^3 N^3 T^6 / 3^7$.  We have 
defined $H_{aa} \equiv \sum_{i=2}^5 H_{x^ix^i} = 4H_{yy}$.

Before looking at the two-point functions, we make two
easy and reassuring observations.  First, the constant term
in the boundary actions is exactly the free energy density, i.e.
minus the pressure, multiplied by the volume, as it should be
\cite{GKT}. 
As an independent check, the pressure can be 
calculated from the entropy $S$ using
the fact that $S = \partial P / \partial T$.  The
entropy for M2- and M5-branes was first determined
in \cite{KlebTseyt}.

The second observation concerns the one-point functions.  
By definition
\be
\langle T^\nu_\mu \rangle = 2 \frac{\delta S_b}{\delta H^0_{\mu\nu}} \ .
\ee
Thus we see that $\langle T^{tt} \rangle = Pd = \epsilon$ and
$\langle T^{ij} \rangle = P\delta^{ij}$ as they should.

The two-point functions are defined similarly:
\be
\langle T^\nu_\mu T^\rho_\lambda \rangle =
-4 \frac{\delta^2 S_b} {\delta H^0_{\mu\nu} \delta H^0_{\lambda\rho} } \ .
\ee
To calculate the two-point functions, we will need to use
our series expansion for $H_K$, (\ref{bulkgravone}) and 
(\ref{bulkgravtwo}), substituting explicit expressions 
for the $H'$ in the boundary action. 
Some details of this calculation have been included as an
appendix.
We find that
the resulting two-point functions match the predictions from 
the Ward identities.  In particular, we find
(\ref{fform}) with $G_1$ given precisely by 
(\ref{Ward1})--(\ref{Wardtwopt}).
%To get precise agreement, we needed to make two
%assumptions about the Green's functions obtained 
%from the Ward identities.  One, the only
%poles in the Green's functions come from 
%sound propagation.  Two, the free parameter
%$n$ in (\ref{fform}) is this thermal factor
%$(\exp(\omega/T)-1)^{-1}$.

\section{Conclusion}

We have succeeded in calculating components of
the stress-energy tensor two-point functions
relevant for sound propagation on 
M2- and M5-branes in the hydrodynamic
limit.  These two-point functions
have poles in the complex
frequency plane in precisely the locations
predicted by hydrodynamics.  Moreover,
the two-point functions calculated
from gravity satisfy the field theory
conformal Ward identities, providing
an elaborate check of the
AdS/CFT correspondence 
at finite temperature.
The agreement between gravity and
field theory gives further support
for the Lorentzian signature prescription
of \cite{SS, HS} for 
calculating correlation functions.

Presumably, a general 
statement
can be made that
correlation functions
calculated from 
Schwarzschild black
holes in $AdS$ of an
arbitrary dimension will
show evidence of sound
propagation.
We have checked
this statement for M2- and M5-branes.
Previously, the statement had been
checked only for D3-branes
\cite{PSSsound}.

To see a familiar phenomenon such as
sound emerge from a
poorly understood theory, such as
those that are meant to 
describe M-branes,
is in the author's mind 
interesting and remarkable.
In the M5-brane case, there
is not even a Lagrangian
description, and yet we can
understand sound waves.

An interesting
direction to pursue is gauge/gravity
duality for non-conformal field theories
in the hydrodynamic limit.
The diffusion constants and viscosities
calculated should
exhibit a scale dependence.
One candidate gauge/gravity duality
where we have high temperature
asymptotics 
is the 
finite temperature 
Klebanov-Tseytlin solution
\cite{GHKT}.  We leave investigation
of these issues for future work.

\section*{Acknowledgments}

I would like to thank 
A.~Starinets especially for some 
very helpful suggestions.
I would also like to thank
D.~T.~Son and 
J.~Walcher for conversations
and correspondence.  Finally,
I would like to thank O.~DeWolfe
and 
N.~Drukker for comments on the manuscript.
This research was supported in 
part by the National Science Foundation under
Grant No. PHY99-07949.

\begin{appendix}

\section{The Bulk Graviton and Correlation Functions}

We present here in more detail some intermediate steps
for calculating stress-energy two-point functions for
M-branes.

By assumption, the bulk graviton components
$H_{xt}$, $H_{tt}$, $H_{s}$, and $H_{yy}$ take on the
values $H^0_{xt}$, $H^0_{tt}$, $H^0_{s}$, and $H^0_{yy}$
at the boundary $u=0$.  However, to calculate the two-point
functions, we also need an expression for $H(u)'$
at the boundary.
% and hopefully this expression
%will cancel the $1/\epsilon^2$ divergence.
Define $P_{\mu\nu} = H_{\mu\nu} Q$
where $Q = 2\omega_2^2 -q_2^2 + \frac{1}{3} i q_2^2 \omega_2$.
Then, for the M2-branes, from the solution to 
(\ref{bulkgravone}), we find
\begin{subequations}
\begin{eqnarray}
\left. P_{tt}'\right|_{u=\epsilon} = 
-\left. P_{s}'\right|_{u=\epsilon}
 &=& \epsilon^2 \left(
3(2\omega_2q_2H_{xt}^0 - q_2^2 H_{tt}^0 + \omega_2^2 H_s^0 )
+ iq_2^2\omega_2H_{yy}^0 \right) + \ldots \\
\left. P_{yy}'\right|_{u=\epsilon} &=& 
\frac{\epsilon^2}{2}
\Bigl(
-3(2\omega_2q_2 H_{xt}^0 - q_2^2 H_{tt}^0
+\omega_2^2 H_s^0)  - 2i\omega_2^2q_2 H_{xt}^0
\\
&&
\; \; \; \; + 2i \omega_2 (\omega_2^2-q_2^2) H_{yy}^0 - 
i \omega_2^3 H_s^0 +i q_2^2 \omega_2 H_{tt}^0
\Bigr) + \ldots \\
\left. P_{xt}'\right|_{u=\epsilon} &=& 
\frac{\epsilon^2}{2} 
\Bigl(
3 q_2 (2 q_2 H_{xt}^0 -2 \omega_2 H_{tt}^0 + \omega_2 H_s^0 ) \\
&&
\; \; \; \;
- 2iq_2^2 \omega_2 H_{xt}^0 -i q_2 \omega_2^2 H_s^0
+ 2iq_2 \omega_2^2 H_{yy}^0
\Bigr) + \ldots 
\end{eqnarray}
\end{subequations}
The ellipses denote higher order terms in $\epsilon$, $\omega_2$,
and $q_2$.  The leading term in $\epsilon$ is precisely of
the form to cancel the leading $1/\epsilon^2$ divergence
in the boundary action (\ref{bryMt}).

For the M5-branes, there is an analogous expression for $H(u)'$.
Defining $P_{\mu\nu} \equiv H_{\mu\nu} Q$, where now
$Q = 5 \omega_5^2 -q_5^2 + \frac{8}{3} i q_5^2 \omega_5$,
one finds from (\ref{bulkgravtwo}) that
\begin{subequations}
\begin{eqnarray}
\left. P_{tt}'\right|_{u=\epsilon} = 
-\left. P_{s}'\right|_{u=\epsilon}
 &=& \epsilon^2 \left(
\frac{15}{2}
(2\omega_5 q_5 H_{xt}^0 - q_5^2 H_{tt}^0 + \omega_5^2 H_s^0 )
+ 20 iq_5^2\omega_5 H_{yy}^0 \right) + \ldots \\
\left. P_{yy}'\right|_{u=\epsilon} &=& 
\epsilon^2
\Bigl(
-\frac{3}{2}
(2\omega_5 q_5 H_{xt}^0 - q_5^2 H_{tt}^0
+\omega_5^2 H_s^0) + 5i \omega_5 (\omega_5^2-q_5^2) H_{yy}^0 \\
&&
\; \; \; \;  - 
i \omega_5^3 H_s^0 +i q_5^2 \omega_5 H_{tt}^0
-2i \omega_5^2 q_5 H_{xt}^0
\Bigr) + \ldots \\
\left. P_{xt}'\right|_{u=\epsilon} &=& 
\epsilon^2 
\Bigl(
\frac{3}{2} 
q_5 (2 q_5 H_{xt}^0 -5 \omega_5 H_{tt}^0 + \omega_5 H_s^0 ) \\
&&
\; \; \; \;
- 8iq_5^2 \omega_5 H_{xt}^0 -4i q_5 \omega_5^2 H_s^0
+ 20iq_5 \omega_5^2 H_{yy}^0
\Bigr) + \ldots 
\end{eqnarray}
\end{subequations}

The next step in calculating the stress-energy two-point
functions is to insert these expressions for $H(u)'$ into
the boundary action
(\ref{bryMt}) or (\ref{bryMf}). 
The terms in the boundary action
that produce the two-point functions can be 
represented schematically as $H^2$ and $HH'$.  The
$H^2$ pieces produce constant contact terms.
The $HH'$ terms, on the other hand, are responsible
for the sound wave pole structure.  

There is an apparent ambiguity how to proceed.
Looking more closely, 
it should not matter at the level of the action whether
the $HH'$ are written as $H(-k,u) H(k,u)'$ or
$H(k,u) H(-k,u)'$.  However, sending $k \to -k$ sends our
solution $H(u) \to H(u)^*$.  

Recall that
according to \cite{HS}, we should be 
substituting $H_K(u) = (1+n) H(u)  -nH(u)^*$ into the
boundary action, and not just $H(u)$.  $H_K(u)$ is
invariant under the transformation $k \to -k$.  
Rather than carry these cumbersome thermal
factors $n$ along, we will adopt the equivalent
procedure of always writing $HH'$ as
$H(-k,u) H(k,u)'$ and using the simpler 
incoming solution $H(u)$.  
%Let $G_1$ be the Green's function
%that results from this procedure. 
Let $S_b$ be the resulting boundary
action.  We can reconstruct $G_F$
by taking functional 
derivatives of $(1+n) S_b - n S_b^*$
instead of just $S_b$.

In this spirit, for the M2-branes,
the $HH'$ pieces of the boundary 
action become
\begin{eqnarray}
\left. S_b \right|_{HH'} &=& 
\frac{P}{8} \left(2\omega_2^2-q_2^2 + \frac{1}{3}iq_2^2\omega_2\right)^{-1}
\Bigl(
12\omega_2 q_2 \left(2 H_{xt}^0 H_{tt}^0- H_{xt}^0 H_{xx}^0
- H_{xt}^0 H_{yy}^0
\right) \nonumber \\
&&
+3\omega_2^2 \left(2 H_{tt}^0 H_{xx}^0
+2H_{tt}^0 H_{yy}^0- (H_{xx}^0)^2
- (H_{yy}^0)^2
-2 H_{xx}^0 H_{yy}^0
\right) \nonumber \\
&&
+3q_2^2 \left( H_{tt}^0 H_{xx}^0
+H_{tt}^0 H_{yy}^0 -4 (H_{xt}^0)^2
-2 (H_{tt}^0)^2
\right) \nonumber \\
&&
+iq_2^2 \omega_2 \left(
3H_{yy}^0 H_{tt}^0 - 2 (H_{yy}^0)^2
+4 (H_{xt}^0)^2 - H_{tt}^0 H_{xx}^0
\right) \nonumber \\
&&
+4i \omega_2^2 q_2 H_{xt}^0 \left(
 H_{xx}^0-H_{yy}^0 
\right)
+i\omega_2^3 \left(
  H_{xx}^0 -H_{yy}^0
\right)^2
\Bigr) \ .
\end{eqnarray}
The quadratic terms in $H$ are tacitly assumed to be
of the form $H(-\omega, -q) H(\omega, q)$.  Also, 
the integral over $\omega$ and $q$ has been suppressed. 

For the M5-branes, the $HH'$ piece of the boundary action
is similarly 
\begin{eqnarray}
\left. S_b \right|_{HH'} &=& 
\frac{P}{4} \left(5\omega_5^2-q_5^2 + \frac{8}{3}iq_5^2\omega_2\right)^{-1}
\Bigl(
12\omega_5 q_5 \left(5 H_{xt}^0 H_{tt}^0- H_{xt}^0 H_{xx}^0
-4 H_{xt}^0 H_{yy}^0
\right) \nonumber \\
&&
+3\omega_5^2 \left(5 H_{tt}^0 H_{xx}^0
+20 H_{tt}^0 H_{yy}^0- (H_{xx}^0)^2
-16 (H_{yy}^0)^2
-8 H_{xx}^0 H_{yy}^0
\right) \nonumber \\
&&
+3q_5^2 \left( H_{tt}^0 H_{xx}^0
+4 H_{tt}^0 H_{yy}^0 -4 (H_{xt}^0)^2
-5 (H_{tt}^0)^2
\right) \nonumber \\
&&
+8iq_5^2 \omega_2 \left(
6 H_{yy}^0 H_{tt}^0 - 5 (H_{yy}^0)^2
+4 (H_{xt}^0)^2 - H_{tt}^0 H_{xx}^0
\right) \nonumber \\
&&
+32i \omega_5^2 q_5 H_{xt}^0 \left(
 H_{xx}^0-H_{yy}^0 
\right)
+8i\omega_5^3 \left(
  H_{xx}^0 -H_{yy}^0
\right)^2
\Bigr) \ .
\end{eqnarray}
Recall that $H_{yy} = H_{x^2x^2}=H_{x^3x^3}=H_{x^4x^4}=H_{x^5x^5}$ 
for these M5-branes.

\end{appendix}

\end{document}